\documentclass[sigconf,natbib=true]{acmart}
\AtBeginDocument{%
  \providecommand\BibTeX{{%
    \normalfont B\kern-0.5em{\scshape i\kern-0.25em b}\kern-0.8em\TeX}}}
\usepackage{multirow}
\usepackage{pifont}
\usepackage{subcaption}
\usepackage{graphicx}
\usepackage{caption}

\newcommand{\cmark}{\ding{51}}%
\begin{document}

\title{CARCA: Context and Attribute-Aware Next-Item Recommendation via Cross-Attention}


\author{Ahmed Rashed}
\email{ahmedrashed@ismll.uni-hildesheim.de}
\orcid{}
\affiliation{%
  \institution{Information Systems and Machine Learning Lab, University of Hildesheim}
  \country{Germany}
}
\author{Shereen Elsayed}
\email{elsayed@ismll.uni-hildesheim.de}
\affiliation{%
  \institution{University of Hildesheim}
  \country{Germany}
}

\author{Lars Schmidt-Thieme}
\email{schmidt-thieme@ismll.uni-hildesheim.de}
\affiliation{%
  \institution{Information Systems and Machine Learning Lab, University of Hildesheim}
  \country{Germany}
}

\renewcommand{\shortauthors}{Rashed, et al.}

\begin{abstract}
In sparse recommender settings, users' context and item attributes play a crucial role in deciding which items to recommend next. Despite that, recent works in sequential and time-aware recommendations usually either ignore both aspects or only consider one of them, limiting their predictive performance. In this paper, we address these limitations by proposing a context and attribute-aware recommender model (CARCA) that can capture the dynamic nature of the user profiles in terms of contextual features and item attributes via dedicated multi-head self-attention blocks that extract profile-level features and predicting item scores. Also, unlike many of the current state-of-the-art sequential item recommendation approaches that use a simple dot-product between the most recent item's latent features and the target items embeddings for scoring, CARCA uses cross-attention between all profile items and the target items to predict their final scores. This cross-attention allows CARCA to harness the correlation between old and recent items in the user profile and their influence on deciding which item to recommend next. Experiments on four real-world recommender system datasets show that the proposed model significantly outperforms all state-of-the-art models in the task of item recommendation and achieving improvements of up to 53\% in Normalized Discounted Cumulative Gain (NDCG) and Hit-Ratio. Results also show that CARCA outperformed several state-of-the-art dedicated image-based recommender systems by merely utilizing image attributes extracted from a pre-trained ResNet50 in a black-box fashion.
\end{abstract}






\maketitle

\section{Introduction}
Nowadays, sequential recommendation systems play an essential role in many online platforms, including but not limited to online shops, online media providers, and social networks. In such platforms, users usually exhibit strong dynamic behavior heavily influenced by ever-changing users' contexts and content information. Due to such dynamic profiles, recent time-aware sequential recommendation approaches that capture sequential patterns in the user profiles \cite{kang2018self, ZhouWZZWZWW20,wu2020sse,wu2019stochastic,sun2019bert4rec} have shown superior performance when compared against traditional non-sequential techniques \cite{rendle2012bpr,song2019autoint} and context-aware models \cite{rendle2010factorization}. However, despite their competitive performance, many of those approaches only utilize the sequential order of consumed items and assume equal time gaps between the interactions. Such an approach completely ignores the actual time gaps between the consumed products and the context in which the user interacted with them. These two aspects encapsulate vital information, which is crucial for determining what item to recommend next. A third aspect that usually gets left out is the item attributes, which are indispensable for any model to overcome highly sparse settings and achieve superior performance \cite{ZhouWZZWZWW20,rashed2019attribute,rashed2020multirec,zhang2019feature}. Additionally, many recent sequential ranking models for item recommendation share a principal limitation as they rely only on the latent features of the most recent item in the user profile to predict scores for the target items. Such a scoring approach significantly downweights the older items' influence on the next items to be recommended.

To tackle these limitations and address the three aspects, we propose a flexible context and attribute-aware recommendation model (CARCA) that captures user profiles' dynamic nature and contextual changes seamlessly alongside any available item attributes. CARCA utilizes several multi-head attention blocks to capture the evolving patterns in the user profile and utilizes a separate dedicated cross-attention block to capture the influence of all previous historical interactions on the target items to be recommended. 

The contributions of this paper can be summarized as follows:

\begin{description}
 \item[$\bullet$] We introduce a versatile context and attribute-aware model for item recommendation (CARCA), which can be applied to diverse settings and can leverage any additional item attributes.
 \item[$\bullet$] We evaluate the proposed model on four real-world datasets for item recommendation. Results show that the proposed CARCA model significantly outperforms all state-of-the-art models in item recommendation and achieves improvements of up to 53\% in Normalized Discounted Cumulative Gain (NDCG) and Hit-Ratio. Results also show that CARCA outperforms several state-of-the-art image-based recommender systems by merely using precomputed image features.
 \item[$\bullet$] 
We conduct a comprehensive ablation study to show the effect of the different model components on the prediction performance.
\end{description}

\section{Related Work}
There has been a plethora of context and attribute-aware recommendation models that have shown consistent competitive performance on a wide variety of recommendation tasks, including, but not limited to, next-item recommendation \cite{rashed2020multirec,sun2019bert4rec,ZhouWZZWZWW20,zhang2019feature,wu2019stochastic}, rating prediction \cite{rashed2019attribute,rendle2010factorization,zhang2017autosvd++}, and click-through rate prediction \cite{guo2017deepfm,zhou2018deep,xin2019cfm,zhou2019deep}. Context-aware models \cite{rendle2010factorization,guo2017deepfm,zhou2018deep,xin2019cfm,zhou2019deep} can achieve superior performance compared with other models due to their ability to capture the high variability of users' behaviors and anticipate their next preferences. On the other hand, attribute-aware models \cite{he2016vbpr,rashed2019attribute,zhang2017autosvd++,rashed2020multirec,ZhouWZZWZWW20,zhang2019feature} utilize the additional item, and user attributes to generate rich latent representations and provide better recommendations.
These additional attributes proved crucial in getting high-quality recommendations in highly sparse settings with rich item attributes such as in online fashion stores \cite{he2016vbpr} and unique item recommendation settings like online auctions \cite{rashed2020multirec}.

\textbf{Context-aware models} can be categorized into two main groups, models that utilize contextual features and other additional attributes in plain single vector format \cite{rendle2010factorization,guo2017deepfm,xin2019cfm} and time-aware models that utilize the time and the sequential order of user's interactions. A popular example that belongs to the first group is the factorization machines (FM) \cite{rendle2010factorization} relying on mining interaction between latent embeddings of the various attributes and contextual features. Such a features extraction approach was later improved by adding deep neural networks \cite{he2017neural,guo2017deepfm} and attention mechanisms \cite{xiao2017attentional} to capture higher-order interactions between the different features' latent vectors. DeepFM \cite{guo2017deepfm} another popular model that utilizes a separate deep neural network to extract non-linear attributes representations along with a traditional FM component. We further shed light on this model in the experiments section, as we propose to use it as a baseline. Similarly, NFM \cite{he2017neural} uses a dedicated deep neural network component on top of the latent features extracted from an FM layer to capture higher-order interactions. Finally, CFM \cite{xin2019cfm} which is one of the latest model in this family, utilized convolutional neural networks on the attributes' latent vectors outer product for better representations learning. However, recent studies \cite{ferrari2020critically} have shown that it performed significantly worse when compared against well tune simple baselines. Nevertheless, despite the competitiveness of such context-aware approaches in item rating prediction and click-through rate prediction tasks, they are significantly outperformed by time-aware sequential models when employed in item recommendation tasks in implicit feedback settings \cite{kang2018self,ZhouWZZWZWW20,zhang2019feature}. Such inferior performance is due to their inability to capture the sequential patterns in the user's historical interactions as they only treat the contextual features as static input features vectors.

The second group of \textbf{context-aware models} are the \textbf{time-aware sequential models} \cite{kang2018self,ma2019hierarchical,ZhouWZZWZWW20,wu2019stochastic,sun2019bert4rec,zhang2019feature,wang2020next,wang2020time,li2020time} which achieve state-of-art performances on item recommendation tasks. Even though these models do not use explicit contextual features, they can capture the evolving user behavior by mining the sequential patterns in their historical interactions. Earlier approaches such as GRU4Rec \cite{hidasi2018recurrent} relied on recurrent neural networks to mine such sequential patterns. This approach was later improved by using bidirectional transformer blocks in BERT4Rec \cite{sun2019bert4rec}. Another recent approach proposed by Kang et al. \cite{kang2018self} that utilizes self-attention blocks to extract the sequential patterns in past user interactions is SASRec. SASRec also utilized a dot-product between the sequential latent features of the most recent item in the user profile and the target items' embeddings for scoring. SASRec was later extended by adding personalized latent user vectors in the current state-of-the-art model SSE-PT \cite{wu2020sse}, adding more robust regularization through stochastic shared embeddings in SSE-SASRec \cite{wu2019stochastic}, adding the ability to model the time intervals between interactions in TiSASRec \cite{li2020time}, and finally adding the ability to handle sparse categorical attributes in the state-of-the-art hybrid S\textsuperscript{3}Rec model \cite{ZhouWZZWZWW20}. All of those approaches, however, maintained the same limited scoring approach that was used in the original SASRec. Lastly, a similar parallel work by Wang et al. \cite{wang2020time} proposed an occasion-aware model (OAR) that utilizes the timestamps embeddings and recurrent memory network to predict the next item to be recommended. OAR also utilized the same dot-product scoring similar to SASRec. We propose using all of the later seven models as baselines in our experimental section as their implementations were readily available. 

Besides context-aware models, \textbf{attribute-aware models} have also shown very competitive performance in various settings despite their inability to capture the users' contexts and their evolving behavior. Earlier approaches such as mSDA-CF \cite{li2015deep} utilized a dedicated denoising auto-encoder component for extracting latent item features from its attributes. This approach was later improved by using a contracted autoencoder in the AutoSVD model \cite{zhang2017autosvd++}. Also, the VPBR model \cite{he2016vbpr} using pre-computed item attributes such as its latent image features has significantly outperformed many non-attribute-based approaches. This approach was later outperformed by a dedicated image-based item recommendation model \cite{ijcai2019650}, which utilizes two CNN networks ensembled to extract global features and local features from different regions of interest. We likewise propose using both approaches as baselines in our experiments. Another recent state-of-the-art attribute-aware model (GraphRec) was proposed by Rashed et al. \cite{rashed2019attribute} that utilizes graph features along with users and items attributes for rating prediction. We further shed light on this model in the experimental section as we propose using it as an additional baseline. 

With motivation set forth from the literature review, we draw the following insight: plenty of works where the user context, time, and item attributes have been actively used to model the user preferences in various recommendation tasks. However, many of these works only focus on one or two of these three aspects, ignoring the correlation between complete contextual information and the crucial item attributes. Also, many of the recent sequential next-item recommendation models rely only on the sequential latent features of the most recent item for scoring, which only captures a limited view of the full user profile. Another essential feature that needs to be considered is the ability to handle not just sparse categorical attributes besides the contextual information but also any number of real-valued attributes to ensure the mode generalizability for different settings. Our work is the first to tackle the all of these aspects and limitations simultaneously for the next-item recommendation task while having the ability to utilize any numerical item attributes.
%%%%%%%%%%%%%%%
\makeatletter
\DeclareRobustCommand*\cal{\@fontswitch\relax\mathcal}
\makeatother
\newcommand\N{{\mathbb N}}
\newcommand\R{{\mathbb R}}
\newcommand\argmax{\arg\,\text{max}}

\section{Problem Definition}
An \textbf{item recommendation problem} consists of
a set ${\cal U}:=\{1,\ldots,$ $U\}$ of \textbf{users}, 
a set ${\cal I}:= \{1,\ldots,I\}$ of \textbf{items} and 
a sequence ${\cal D}:= ((u_1,i_1),\ldots,(u_N,i_N))\in({\cal U}\times{\cal I})^*$
of their past \textbf{interactions} originating from an unknown distribution $p$
on user/item pairs (with $U,I,N\in\N$).
Sought is a model $\hat p: {\cal U}\rightarrow(\R^+_0)^{\cal I}$
for the unknown conditional density $p(i\mid u)$, i.e., given
a loss function $\mathcal{L}:{\cal I}\times (\R^+_0)^{\cal I}\rightarrow\R$
with minimal expected loss
\begin{align*}
    {\mathbb E}_{(u,i)\sim p}\ \mathcal{L}(i, \hat p(u))
\end{align*}

One calls the problem \textbf{having item attributes}, if
additionally there is a given matrix $A^{\text{IT}}\in\R^{I\times j}$ containing for item an attribute vector with $j\in\N$ attribute values each.

One calls the problem \textbf{having context}, if each \textsl{interaction} has additional
attributes, i.e., 
  ${\cal D}:= ((u_1,i_1,c_1),\ldots,(u_N,i_N,c_N))\in({\cal U}\times{\cal I}\times\R^{l})^*$
(with $l\in\N$)
is a sample from an unknown distribution on user/item/context triples
and the goal is to find
a model $\hat p: {\cal U}\times\R^{l}\rightarrow(\R^+_0)^{\cal I}$
for the unknown conditional density $p(i\mid u,c)$, i.e., given
a loss function $\mathcal{L}$ with minimal expected loss
\begin{align*}
    {\mathbb E}_{(u,i,c)\sim p}\ \mathcal{L}(i, \hat p(u,c))
\end{align*}
The most frequently encountered context is an absolute \textbf{time-stamp} at which
the user interacted with an item (for example measured as a real number in Unix Time).

Many papers simplify the problem, dropping the sequential nature of the problem.
Consequently, they model the data as a \textsl{set} of user/item interactions
(instead of as a sequence) and also evaluate on \textsl{non-sequential splits}, i.e., take out items
at random positions (instead of only at the end). While this allows simpler models,
it has the significant disadvantage that these splits cannot occur in reality. We will therefore
look only at the sequential problem as stated above.

Sequential approaches on the other hand usually consider all users to have profiles $P^{u}_{t}$ that contain the sequence of their previously interacted items $P^{u}_{t} := \{i^{P}_{1}, i^{P}_{2}, ..., i^{P}_{|P^{u}_{t}|}\}$ along with their attributes $A^{u}_{t} \in \mathbb{R}^{ |P^{u}_{t}| \times j} $ and their interactions' contextual features $C^{u}_{t} \in \mathbb{R}^{ |P^{u}_{t}| \times l} $ such as timestamps. The main goal of the sequential item recommendation task will be to rank a target list of items $O^{u}_{t+1} := \{i^{O}_{1}, i^{O}_{2}, ..., i^{O}_{|O^{u}_{t+1}|}\}$ based on their likelihood of being interacted with by the target user $u$ at time $t+1$ while similarly considering their attributes and contextual features existing at that time point.

%%%%%%%%%%%%%%%

% \section{Problem Definition}
% In implicit feedback recommendation settings, there exist a set of items $I$, a set of users $U$ with their profiles $P^{u}_{t}$. These profiles contain the sequence of their previously interacted items $P^{u}_{t} := \{i^{p}_{1}, i^{p}_{2}, ..., i^{p}_{|P^{u}_{t}|}\}$ along with their attributes $A^{u}_{t} \in \mathbb{R}^{ |P^{u}_{t}| \times j} $ and their interactions' contextual features $C^{u}_{t} \in \mathbb{R}^{ |P^{u}_{t}| \times l} $ such as timestamps. The main goal of the item recommendation task is to rank a target list of items $O^{u}_{t+1} := \{i^{o}_{1}, i^{o}_{2}, ..., i^{o}_{|O^{u}_{t+1}|}\}$ based on their likelihood of being interacted with by the target user $u$ at time $t+1$ while similarly considering their attributes and contextual features existing at that time point.

\section{Methodology}
To capture the evolving users' behaviors existing in their profiles $P^{u}_{t}$, the proposed CARCA model utilizes two analogous multi-head self-attention based branches. The left branch is a series of self-attention blocks that extract the user's profile's contextual information and item features. On the other hand, the right branch consists of a multi-head cross-attention block that captures the influence of the left branch's profile-level features on the target items $O^{u}_{t+1}$ while taking into consideration the target items' attributes and contextual features. The second branch is also responsible for generating the ranking score for each target item in $O^{u}_{t+1}$.

\begin{figure*}[!ht]
\centering
\includegraphics[scale=0.35]{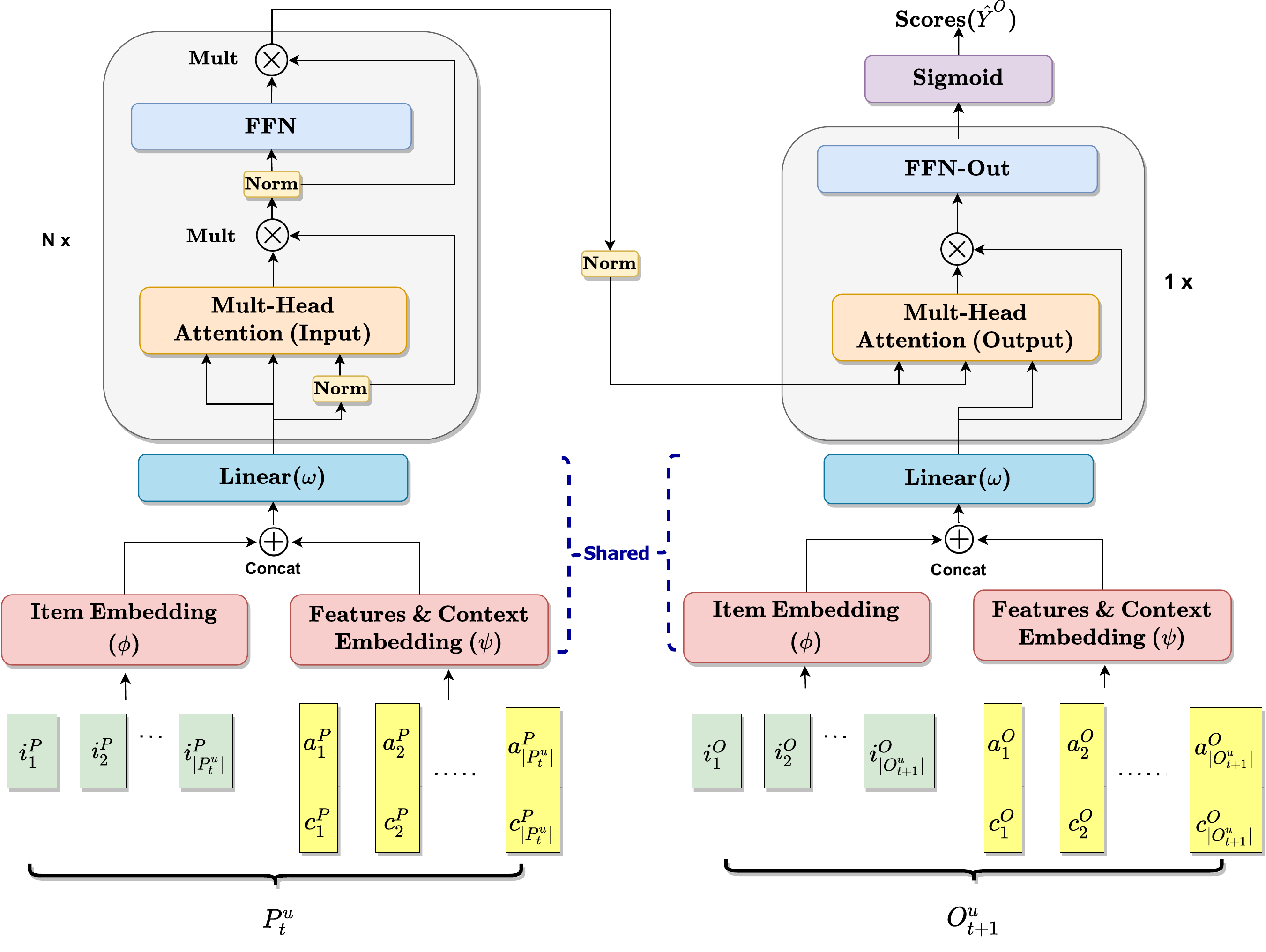}
\caption{Illustration of the CARCA model, which is composed of two main branches, namely the profile-level features extraction branch on the left and the target items cross-attention scoring branch on the right.}
\label{Arch}
\end{figure*}

Figure \ref{Arch} illustrates the architecture of the CARCA model, which will be discussed in detail in the following subsections.

\subsection{Embedding Layers}
The first part of the features extraction pipeline in both branches is the embedding functions that will extract the initial items latent features to be fed to the self-attention blocks. To achieve that, we utilize two separate dedicated embedding functions $\phi$ and $\psi$. The first embedding function $\phi: \mathbb{R}^{I} \rightarrow \mathbb{R}^{d}$ is used to extract the first half of the item's latent features $z_{i} \in \mathbb{R}^{d}, i \in P^{u}_{t} \cup O^{u}_{t+1} $ from the item's one-hot encoded vectors $x_{i} \in \mathbb{R}^{I}$. The second function $\psi: \mathbb{R}^{j+l} \rightarrow \mathbb{R}^{g}$ extracts the second half of the latent features $q_{i} \in \mathbb{R}^{g}$ from the item's contextual features $c_{i} \in \mathbb{R}^{l}$ and attributes $a_{i} \in \mathbb{R}^{j}$. After extracting the two partial latent feature vectors, both of them are concatenated and fed into a third embedding layer $\omega: \mathbb{R}^{g+d} \rightarrow \mathbb{R}^{d}$ to generate the final item's latent features $e_i \in \mathbb{R}^{d}$ as follows:

\begin{equation} 
z_{i}=\phi(x_{i})= x_{i}W^{\phi} + b^{\phi}, \: \: W^{\phi} \in \R^{I\times d}, \:  b^{\phi} \in \R^{d} 
\end{equation} 

\begin{equation}
q_{i}= \psi(a_{i},c_{i}) =  \text{concat}_{col}(a_{i},c_{i})W^{\psi} + b^{\psi}, \: \: \: W^{\psi} \in \R^{ (j+l)\times g}, \: b^{\psi} \in \R^{g} 
\end{equation} 

\begin{equation} 
 e_{i}= \omega(z_{i},q_{i}) = \text{concat}_{col}(z_{i},q_{i})W^{\omega} + b^{\omega},\: \: \: W^{\omega} \in \R^{ (g+d)\times d},\: \: \: b^{\omega} \in \R^{d} 
\end{equation}

\noindent where $W^{\phi}$, $W^{\psi}$, $W^{\omega}$ are the weight matrices of the embedding functions, and $b^{\phi}$, $b^{\psi}$, $b^{\omega}$ are their bias vectors. $\text{concat}_{col}$ represents column-wise concatenation of vectors.

Finally, since both the user profile $P^{u}_{t}$ and the target items $O^{u}_{t+1}$ have the same structure and format, we utilize the same embedding pipeline on both of them while sharing the network's weights. It is worthy to note that we also tried to use a single embedding layer for all input features concatenated into one long vector. However, the performance was inferior to this setup.

\subsection{Self-Attention Blocks}

After extracting the embeddings of all items in the user profile $E^{P} := \{e^{P}_{1}, e^{P}_{2}, ..., e^{P}_{|P^{u}_{t}|}\}$ and the embeddings of the target items $E^{O} := \{e^{O}_{1}, e^{O}_{2}, ..., e^{O}_{|O^{u}_{t+1}|}\}$, we use two separate self-attention components to extract profile-level features and ranking the target items.

\subsubsection{Profile-Level Self-Attention Blocks}
To extract the profile-level features, we feed the item embeddings $E^{P}$ into a series of multi-head self-attention blocks. We first feed the item embeddings into the first part of a block, which is a multi-head self-attention layer that utilizes the scaled dot product. \cite{vaswani2017attention}

\begin{equation} 
\textrm{Attention}(\textbf{Q}, \textbf{K}, \textbf{V}) = \textrm{softmax}\left( \frac{\textbf{Q} \textbf{K}^{T}}{\sqrt{\frac{d}{H}}}\right) \textbf{V}
\end{equation} 

\begin{equation} 
\begin{split}
S^{P} &= \textrm{ SA}(\textbf{E}^{P}) \\& = \text{concat}_{col} \left( \textrm{Attention}(\textbf{E}^{P}\textbf{W}^{Q}_{h}, \textbf{E}^{P}\textbf{W}^{K}_{h}, \textbf{E}^{P}\textbf{W}^{V}_{h}) \right)_{h=1:H}
\end{split}
\end{equation}

\noindent where $\textbf{Q}$, $\textbf{K}$ and $\textbf{V}$ represents the queries, keys and values respectively. $\textbf{W}^{Q}_{h}$, $\textbf{W}^{K}_{h}$, $\textbf{W}^{V}_{h} \in \mathbb{R}^{d \times \frac{d}{H}}$ represent the linear projection matrices of the head at index $h$ and $H$ represents the total number of heads in the attention block. Additionally, $\text{concat}_{col}$ concatenates vectors column-wise and $\sqrt{\frac{d}{H}}$ is a scaling factor that controls the values of the inner products. 

The next step is to feed the output of the self-attention layer into the second part, which is a point-wise two-layer feed-forward network that is applied identically to all elements $S^{P}_r$ of $S^{P}$ with sharing parameters similar to the original transformers \cite{vaswani2017attention}

\begin{equation} 
\begin{split}
F^{P} &= \textrm{ FFN}(S^{P}) \\& = \text{concat}_{row} \left( \textrm{Leaky\_ReLU} (S^{P}_r W^{(1)} + b^{(1)})W^{(2)} + b^{(2)}  \right)_{r=1:|P^{u}_{t}|}
\end{split}
\end{equation}

\noindent where $W^{(1)}$, $W^{(2)} \in \mathbb{R}^{d \times d}$ are the weight matrices of the two feed-forward layers, and $b^{(1)}$, $b^{(2)} \in \mathbb{R}^{d}$ are their bias vectors. Additionally, $\text{concat}_{row}$ concatenates vectors row-wise. These feed-forward networks introduce essential non-linearity, which allows us to capture higher-order interactions between the features.

Finally, to capture more expressive profile-level features, we stack a series of those self-attention blocks where the $b$-th ($b > 1$) blocks are defined as follows:

\begin{equation} 
S^{P,(b)} = \textrm{SA}(\textbf{F}^{P,(b-1)}) 
\end{equation} 

\begin{equation} 
F^{P,(b)} = \textrm{FFN}(S^{P,(b)}) 
\end{equation} 

\noindent where the 1-st block is defined as $S^{P,(1)} = S^{P}$ and $F^{P,(1)} = F^{P}$ 
Inspired by \cite{vaswani2017attention,kang2018self}, we utilized residual connections \cite{he2016deep}, layer normalization \cite{ba2016layer} and dropout \cite{srivastava2014dropout} to alleviate the problems of overfitting and instability. One slight difference, though, is that we used multiplicative residual connections instead of additive ones as they provided better performance and CARCA's architecture is relatively not very deep compared to the ResNet models. We also omitted the causality constrain to allow bidirectional representation learning similar to BERT\cite{devlin2019bert} and S\textsuperscript{3}Rec \cite{ZhouWZZWZWW20} models. It is also important to note that we omitted the positional encoding since this information already exists explicitly in the contextual features such as the interaction's timestamp.

\subsubsection{Target-Level Cross-Attention Layer}
Following the original transformers architecture and inspired by recent click-through-rating prediction models \cite{zhou2019deep}, and in contrast to many of the recent ranking-based sequential and context-aware item recommendation approaches \cite{kang2018self,ZhouWZZWZWW20,wu2020sse,wang2020time,wang2020next}, we utilize a dedicated multi-head cross-attention block to predict the likelihood scores of the target items. The dedicated block aims to capture the interaction between all profile-level features $F^{P,(b)}$ and the targets items instead of the widely used approach of only using the dot product between the last most recent element in $F^{P,(b)} $ and the target items initial embeddings $E^{O}$. We further shed light on the comparison between these two approaches in the experimental section.

To calculate the scores $\hat{Y}^{O} \in \mathbb{R}^{|O^{u}_{t+1}|} $ of the target items $O^{u}_{t+1}$, we feed their embeddings in a multi-head cross-attention block as the query input while using the normalized profile features $F^{P,(b)}$ from the left branch's last self-attention block as the keys and values

\begin{equation} 
\begin{split}
S^{O} &= \textrm{ CA}(\textbf{E}^{O},\textbf{F}^{P,(b)}) \\&=  \text{concat}_{col} \left(\textrm{Attention}(\textbf{E}^{O}\textbf{W}^{Q}_{h}, \textbf{F}^{P,(b)}\textbf{W}^{K}_{h}, \textbf{F}^{P,(b)}\textbf{W}^{V}_{h})\right)_{h=1:H}
\end{split}
\end{equation} 

\begin{equation} 
\hat{Y}^{O} = \textrm{FFN-Out}(S^{O}) = \text{concat}_{row} \left( \sigma(S^{O}_r W^{O} + b^{O}) \right)_{r=1:|O^{u}_{t+1}|}
\end{equation} 

\noindent where $\sigma$ is the sigmoid activation. $W^{O} \in \mathbb{R}^{d \times 1}$ and $b^{O} \in \mathbb{R}$ are the weight matrices and bias vector of the output layer. 

Such a setup allows us to generate different latent representations for the target items that capture the interactions between the full profile-level features and their embeddings. It also allows us to have an arbitrarily sized list of target items as an input $O^{u}_{t+1}$. It is also worth noting that we omitted the initial multi-head self-attention block that exists on top of the original transformers architecture's output tokens because the target items in our settings are scored independently. However, this component might be useful in other scenarios where the interactions between the target items are relevant to the prediction task, such as in the next basket recommendation task.

\subsection{Optimizing CARCA}
To simplify the training process, we followed the same training protocol proposed by SASRec and TiSASRec \cite{kang2018self, li2020time}. For each user we exclude his last interaction and we convert the user profile sequence into a fixed-length input list of items $P^{u} =\{i^{P}_{1}, i^{P}_{2}, ..., i^{P}_{|P^{u}_{t}|-1}\}$ via truncation or padding. On the other hand, the list of target items is constructed by combining a list of positive items $O^{u(+)}$ and another list of negative items $O^{u(-)}$ with equal length. The positive items list is constructed by right shifting the input list $P^{u}$ to include the user's last interaction $O^{u(+)} =\{i^{P}_{2}, i^{P}_{3}, ..., i^{P}_{|P^{u}_{t}|}\}$ while the negative items list is generated by selecting random negative items $i \notin P^{u}$ and they are given the same contextual features as their corresponding positive ones. It worth noting that we also tried using the last interaction as the only positive target in $O^{u(+)} =\{i^{P}_{|P^{u}_{t}|}\}$, however, the performance was inferior to the list-wise setup. We further shed the light on the comparison between the two splitting approaches in Section 5.5.

Finally, we optimize the CARCA model by minimizing the binary cross-entropy loss using an ADAM optimizer, and the padded items are masked to prevent them from contributing to the loss function.

\begin{equation}
    \mathcal{L}=- \sum_{u \in U} \sum_{r \in O^{u(+)} \cup O^{u(-)}} 
    \left(Y^{O}_r \log(\hat{Y}^{O}_r)+ (1-Y^{O}_r)\log(1-\hat{Y}^{O}_r ) \right)
\end{equation}

%%%%%%%%%%%%%%%%%%%%%%%%%%%%%%%%5

\begin{table}[!ht]
\caption{Datasets Statistics}
\label{datasets}
\small
\begin{center}
  \begin{tabular}{lcccccc}
    \toprule
    Dataset&Users&Items&Interactions&Item Attributes\\
    \midrule
Men & 34,244 & 110,636 & 254,870 & 2048 \\
Fashion & 45,184 & 166,270 & 358,003 & 2048  \\
Games & 31,013 & 23,715 & 287,107 & 506   \\
Beauty & 52,204 & 57,289 & 394,908 & 6507  \\
   \bottomrule
  \end{tabular}
  \end{center}
\end{table}
%%%%%%%%%%%%%%%%%%%%%%%%%%%%%%%%%
\section{Experiments} 
In this section, we conduct multiple experiments to evaluate the performance of CARCA and to answer the following research questions.
\begin{description}
 \item[RQ1] How well does CARCA perform compared to the state-of-the-art recommender system models on item recommendation tasks?
 
  \item[RQ2] How well does CARCA perform compared to the state-of-the-art image-based recommender system models while only using pre-computed item's image attributes?
 
  \item[RQ3] What is the impact of adding the item's attributes and contextual features?
  
  \item[RQ4] What are the impacts of the different components and design choices of the CARCA architecture?
 
\end{description}

\subsection{Datasets}
To evaluate the performance of CARCA and compare its performance against published results, we used the following four diverse and widely used real-world datasets extracted from products' reviews crawled from Amazon.com \cite{kang2018self,wu2020sse,steck2019embarrassingly,ZhouWZZWZWW20,he2016vbpr,ijcai2019650}. All datasets' contextual features were extracted from the interactions' timestamps (i.e., day, month, year, day of week, day of year and week). 
\begin{enumerate}
    \item \textbf{Men} \cite{he2016vbpr}: This dataset contains all items that belong to men’s clothing including all subcategories (gloves, scarves, sunglasses, etc.). The item attributes in this dataset are image based features extracted by using the output of the last fully connected layer of a pre-trained ResNet50 \cite{he2016deep} on the ImageNet dataset \cite{imagenet_cvpr09}.
    
    \item \textbf{Fashion}\cite{ijcai2019650,he2016vbpr}: This dataset contains six fashion categories (men/women’s tops, bottoms and shoes). This dataset's item attributes are also image-based features extracted by the same pre-trained ResNet50 as the Men dataset.
    
    \item \textbf{Games} \cite{kang2018self,wu2020sse}: This dataset contains all items that belong to video games category. The item attributes in this dataset are the price, fine-grained categories, and the item's brand. Most of the items' attributes are discrete and categorical in contrast to the first two datasets. 

    \item \textbf{Beauty} \cite{kang2018self,wu2020sse,ZhouWZZWZWW20}: This dataset contains all items that belong to beauty category. The item attributes in this dataset are the fine-grained categories and the item's brand. All of the items' attributes are also discrete and categorical. 
     
\end{enumerate}

To have a fair comparison against S\textsuperscript{3}Rec \cite{ZhouWZZWZWW20} that uses only categorical attributes, we discretized all real-valued input features (attributes) into different levels. The number of levels was selected as the maximum number after which the S\textsuperscript{3}Rec fails to run because of exceeding the total available memory of the GPU (1080Ti: 11 GB) or the system RAM (64GB). According to this, the maximum number of the selected equally spaced discretization levels were 10, 10, and 50 for the Men, Fashion, and Games datasets, respectively. 
Table \ref{datasets} presents a summary of the most important statistics of the datasets.

%%%%%%%%%%%%%%%%%%%%%%%%%%%%%%%%%%%%%%%%%%%

\subsection{Performance comparison with state-of-the-art item recommendation models (RQ1)}
In this section, we compare the performance of the CARCA model against multiple state-of-the-art ranking-based next-item recommendation models with different capabilities. 
%A short summary of all baselines and their capabilities is indicated in table [].

%%%%%%%%%%%%%%%%%%%%%%%%%%%%%%%%%%%%
\begin{table*}[!ht]
\setlength{\tabcolsep}{2.5pt}
\caption{Performance comparison of the CARCA against state-of-the-art sequential (SEQ), context (CXT) and attribute-aware (ATT) recommendation models.}
\label{Comp1}
\small
\begin{center}
\begin{tabular}{l|ccc|cccccccc}
    \toprule
 &&&& \multicolumn{2}{c}{Men} & \multicolumn{2}{c}{Fashion} & \multicolumn{2}{c}{Games} & \multicolumn{2}{c}{Beauty} \\
 
Model&ATT&CXT&SEQ & HR@10 & NDCG@10  & HR@10 & NDCG@10    & HR@10& NDCG@10   & HR@10 & NDCG@10   \\
\midrule

Random &&& & 0.098 & 0.044 & 0.099 & 0.045 & 0.100& 0.045& 0.099 & 0.045\\
TopPop &&& & 0.415 & 0.269 & 0.407 & 0.262 & 0.519& 0.314& 0.451 & 0.261\\
EASE  \cite{steck2019embarrassingly}&&&& 0.193 & 0.133 & 0.213 & 0.146 & 0.623& 0.465& 0.299 & 0.222\\
GraphRec \cite{rashed2019attribute}&\cmark&&    & 0.374 & 0.219 & 0.419 & 0.244 & 0.613& 0.400& 0.435& 0.273   \\
DeepFM \cite{guo2017deepfm}&\cmark&\cmark& & 0.334 & 0.237 & 0.283 & 0.185 & 0.736& 0.494& 0.464& 0.266   \\
SASRec \cite{kang2018self}&&&\cmark     & 0.397 & 0.259 & 0.381 & 0.245 & 0.742& 0.541& 0.485 & 0.322\\
OAR \cite{wang2020time}&&\cmark&\cmark     & 
0.355&	0.225&	0.340&	0.214&	0.704&	0.496&	0.485&	0.329\\
TiSASRec \cite{li2020time}&&\cmark&\cmark     & 
0.333 & 0.194	& 0.384 &	0.234 & 0.748&	0.533&	 0.492& 0.333\\
BERT4Rec \cite{sun2019bert4rec}&&&\cmark & 0.315 & 0.193 & 0.328 & 0.209 & 0.705& 0.509& 0.478 & 0.318\\
SSE-SASRec \cite{wu2019stochastic}&&&\cmark & 0.397 & 0.257 & 0.385 & 0.248 & 0.754& 0.549& 0.481 & 0.330\\
SSE-PT    \cite{wu2020sse}&&&\cmark  & 0.397 & 0.258 & 0.381 & 0.246 & 0.748(0.775)& 0.545(0.566)& 0.443(0.502) & 0.302(0.337)\\
S\textsuperscript{3}Rec  \cite{ZhouWZZWZWW20}&\cmark&&\cmark     & 0.365 & 0.238 & 0.367 & 0.239 & \underline{0.765}& \underline{0.549}& 0.538 & \underline{0.371}\\
\midrule
SASRec++ (Our extension)&\cmark &\cmark &\cmark    &  \underline{0.500} & \underline{0.315} & \underline{0.546} & \underline{0.344} & 0.752& 0.533& \underline{0.545} & 0.351\\
\midrule
CARCA  (w/o CA) (Ours) &\cmark &\cmark &\cmark   &0.521    & 	0.322 &   0.568	    &   0.359&  0.738	&   0.517   &   0.556 &    	0.358\\
CARCA  (Ours)&\cmark &\cmark &\cmark   & \textbf{0.550*}&\textbf{ 0.349*} & \textbf{0.591*} &\textbf{ 0.381*} &\textbf{0.782*} & \textbf{0.573*}& \textbf{0.579*} & \textbf{0.396}\\ 
   \bottomrule
 \bottomrule
\multicolumn{1}{l}{Improv. vs best published baseline (\%)}& && &38.65 & 35.87 & 53.71 & 53.24 & 2.20 & 4.38 & 7.70 &	6.74 \\ 
\multicolumn{1}{l}{Improv. vs SASRec++ (\%)} & && &10.09&10.79&8.25&10.67&3.96&7.64&6.31&12.95 \\
   \bottomrule
\end{tabular}
  \\ \small (*) Significantly outperforms the best baseline at the 0.01 levels. 
  \\ Published results of SSE-PT are indicated in parentheses. 
  \end{center}
\end{table*}
%%%%%%%%%%%%%%%%%%%%%%%%%%%%%%%%%%%%%

\subsubsection{Evaluation Protocol}
To evaluate CARCA and other baseline models' performance, we used the widely adopted leave-one-out protocol \cite{kang2018self,wu2020sse,ZhouWZZWZWW20,he2016vbpr,ijcai2019650,wang2020time}. In this protocol, the two last interactions of each user are held out for validation and test while the rest of the interactions are used for training. To evaluate the model's performance, we sample 100 negative items that were not interacted with by the user, give them all the same context as their corresponding positive test item, and rank the positive test item among them.
Lastly, for each user, we truncate the ranked list at a threshold value of 10. We measure the overall quality using the average Hit-Ratio (HR) and the Normalized Discounted Cumulative Gain (NDCG) across users.

To ensure the statistical significance of the reported results, we report the average metrics across five different runs on the test set, each with a different set of random negative items, and we used a paired t-test for measuring the significance. The hyper-parameters of all models were tuned on the validation set using grid search. We also tried the best hyper-parameters reported in the baselines' original papers if they were available.

\subsubsection{Models}

\begin{enumerate}
 \item \textbf{Random}: A simple baseline model that ranks items randomly.
 \item \textbf{TopPopular}: A naive baseline model that ranks items based on their popularity.
 \item \textbf{EASE} \cite{steck2019embarrassingly}: A shallow auto-encoder based model that utilize the closed-form solution of the Frobenius norm objective function in highly sparse settings.
\item \textbf{GraphRec} \cite{rashed2019attribute}: This is an extended version of the state-of-the-art attribute-aware GraphRec model for item recommendation in implicit feedback settings. We replaced the means squared error (MSE) loss function with a logistic loss function, and we trained the model by sampling positive and negative items. 
 \item \textbf{DeepFM} \cite{guo2017deepfm}: A widely used model for click-through rate prediction that relies on learning high order feature interactions using an ensemble of Factorization Machines and deep neural networks. We modified this model for next-item recommendation by optimizing it using negative sampling and minimizing a logistic loss function.
 
 \item \textbf{SASRec} \cite{kang2018self}: A state-of-the-art sequential recommendation model that utilizes self-attention blocks to predict the next item to be recommended. It also uses the dot-product between the most recent item's sequential latent features and the target item's embeddings as the scoring function.
  \item \textbf{TiSASRec} \cite{li2020time}: An improved version of SASRec that utilizes time-aware positional embeddings for modeling time intervals between interactions.
 \item \textbf{OAR} \cite{wang2020time}: A state-of-the-art sequential recommendation model that utilizes recurrent memory networks and gating layers. This model also utilizes the timestamp as contextual features.
 \item \textbf{SSE-SASRec }\cite{wu2019stochastic}: An improved version of SASRec that utilizes stochastic shared embeddings for regularization.
 \item \textbf{BERT4Rec }\cite{sun2019bert4rec}:  A state-of-the-art model that utlizes bidirectional transformers for next item recommendations.
 
 \item \textbf{SSE-PT} \cite{wu2020sse}: A state-of-the-art extended version of SASRec that utilizes the latent user vectors along with their historical interactions. 
 \item \textbf{S\textsuperscript{3}Rec} \cite{ZhouWZZWZWW20}: A state-of-the-art attribute-aware sequential recommendation model that utilizes multiple self-supervised mutual information maximization loss components for better representation learning of item attributes. This model can handle categorical attributes only, and real-valued attributes will require discretization in order to be applicable.
 \item \textbf{SASRec++}\textbf{(Ours)} \cite{kang2018self}: This is our context and attribute-aware extended version of SASRec. We used the same initial feature extraction pipeline used by CARCA, and we replaced the one-hot encoded input vectors of SASRec with the extracted items' embedding vectors.
 
  \item \textbf{CARCA (w/o CA)}\textbf{ (Ours)}: A version of our proposed model CARCA that does not utilize the cross-attention component and only utilize the dot-product scoring function between the sequential features of the most recent item and target items similar to other sequential recommendation models.
  
 \item \textbf{CARCA} \textbf{(Ours)}: This is our proposed method with cross-attention between all profile-level features and the target items' embeddings.

\end{enumerate}

It is worth noting that we could not reproduce the original results of the SSE-PT model on the Beauty and Games datasets, although we used the authors' implementation because best hyper-parameters were neither mentioned in the paper nor the code. The authors are also no longer able to recover the best hyper-parameters as per the following GitHub issue \footnote{https://github.com/wuliwei9278/SSE-PT/issues/1} that other community members raised. To mitigate this issue, we report both our results and published results on those respective datasets.

\subsubsection{Results}
Results in Table \ref{Comp1} show that CARCA with cross attention significantly outperforms all state-of-the-art context, sequential and attribute-aware models on various settings and with different item attribute types. CARCA is also able to achieve improvements up to 53\% when compared against SSE-PT as it utilizes the full pre-computed item's image features on the Men and Fashion datasets without needing any data discretization, unlike the S\textsuperscript{3}Rec that only handle categorical attributes. Additionally, CARCA was also able to significantly outperform S\textsuperscript{3}Rec in speed ($\approx$ ~22 times faster) and accuracy on settings with categorical attributes such as the Games and Beauty datasets despite the competitive performance of S\textsuperscript{3}Rec on them. 

Results also show that the CARCA without the cross-attention branch, which utilizes the same dot-product scoring approach adopted by many recent sequential models, can outperform all baselines, including the SASRec++ extended version on three out of the four datasets. However, this version is still outperformed by the cross-attention version, which provides a further lift of 2\% to 6\% in NDCG and HitRatio. This difference between the two CARCA versions shows that older items in the user profile have a significant influence on the next items to be interacted with, and such influence should not be ignored.

One interesting finding is that S\textsuperscript{3}Rec was found to be slightly inferior to SASRec on the Men and Fashion datasets. This is mainly because of the features discretization step that S\textsuperscript{3}Rec needs which leads to a negative impact on its performance due to the unavoidable information loss. On the other hand, this also explains why S\textsuperscript{3}Rec is superior on the Games and Beauty datasets, which has mostly categorical features. Another interesting finding is that BERT4Rec was also inferior to SASRec although it is a more recent approach. These findings match similar results in recently published comparative studies \cite{ZhouWZZWZWW20}.

Finally, it is worth mentioning that the TopPopular model was found to be very competitive on the Men, Fashion, and Beauty datasets due to unevenly distributed items popularities in these datasets similar to the Epinion dataset in the famous study by \cite{dacrema2019we}. We conducted a similar analysis on our test sets items, and we found out that the Gini indexes of the item's popularity in the Fashion and Men datasets (Gini indexes =  0.71 and 0.72) are significantly high, which explains the competitive performance of TopPopular model. Also, the Gini index on Beauty is slightly higher than Games (Beauty = 0.67 and Games = 0.63), which explains why the TopPopular achieved a very competitive performance on that dataset compared to its performance on the games dataset.

\subsection{Performance comparison with state-of-the-art dedicated image-based item recommendation models (RQ2)}
In this section, we compare the performance of the CARCA model against multiple state-of-the-art image-based item recommendation models on the Fashion dataset. Regarding the evaluation protocol, we used the same as the one proposed by SAERS \cite{ijcai2019650} with sampling 500 negative items instead of 100 and using the area under the curve (AUC) score instead of the HitRatio. This allows us to compare CARCA's performance against their published results directly.

\subsubsection{Baselines}
\begin{enumerate}
 \item \textbf{VBPR} \cite{he2016vbpr}: Image-based attribute aware model that utilizes a pre-trained CaffeNet network for extracting item's image features and a BPR model for item recommendation
 \item \textbf{JRL} \cite{zhang2017joint}: A State-of-the-art neural recommender system that leverages the item's image features for Top-N recommendation.
  \item \textbf{SAERS} \cite{ijcai2019650}: State-of-the-art image-based item recommendation model that utilizes two CNN networks ensembled to extract global and local image features from different regions of interest.
\end{enumerate}

\subsubsection{Results}
Results in Table \ref{Comp2} shows that CARCA is able to outperform SAERS and all other image-based models while merely utilizing pre-computed features and without any fine-tuned image-based deep neural network components. These findings show that even fine-tuned image-based features are not sufficient for capturing user interests, but we also need to include the contextual features of the interactions.
%%%%%%%%%%%%%%%%%%%%%%%%%%%%%%%%%%%%%%%%%%%%%%%%%%%
\begin{table}[!ht]
\caption{Performance comparison of the CARCA against state-of-the-art image-based models on the Fashion dataset.}
\label{Comp2}
\small
\begin{center}
\begin{tabular}{lcc}
\toprule

                       Model              & NDCG@10 & AUC   \\
\midrule
Random                                 & 0.012   & 0.501 \\
TopPop                               & 0.125   & 0.595 \\
VBPR \cite{he2016vbpr}                                 & 0.085 & 0.771     \\
JRL  \cite{zhang2017joint}                                & 0.127   & 0.771 \\
SAERS \cite{ijcai2019650}                               & \underline{0.171}   & \underline{0.816} \\
\midrule
CARCA                                  & \textbf{0.184}   & \textbf{0.841} \\
\midrule
\midrule
Improvement over best baseline  (\%) & 7.54    & 3.03 \\
\bottomrule
\end{tabular}
  \end{center}
\end{table}
%%%%%%%%%%%%%%%%%%%%%%%%%%%%%%%%%%%%%%%%%%%%%%%%%%%%
\begin{figure}[!ht]
\centering
\begin{subfigure}[t]{.235\textwidth}
  \includegraphics[trim={3.1cm 0 0 0},clip,scale=0.107]{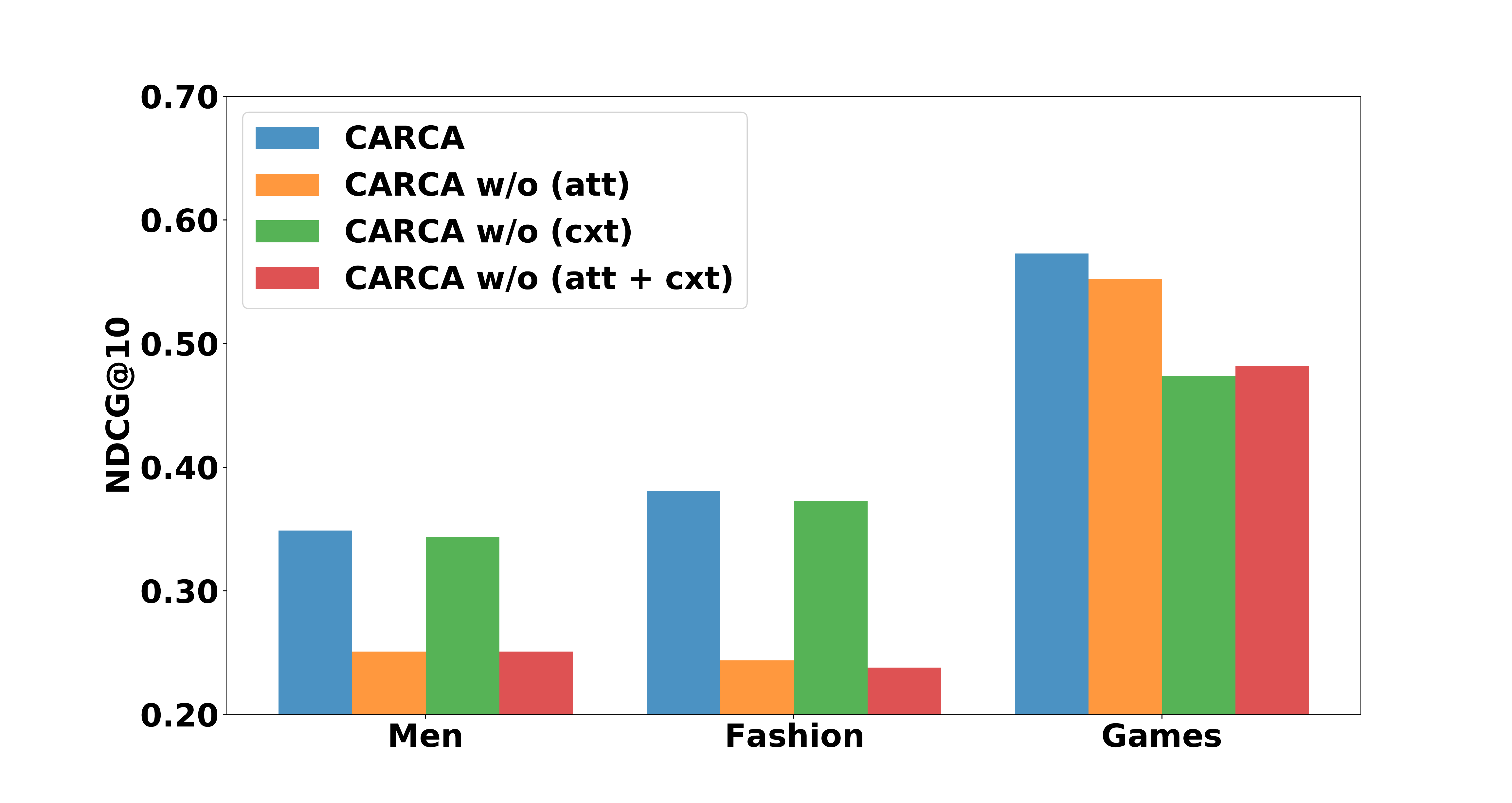}
  \caption{NDCG}
  \label{sub1}
\end{subfigure}
\centering
\begin{subfigure}[t]{.235\textwidth}
  \includegraphics[trim={3.1cm 0 0 0},clip,scale=0.107]{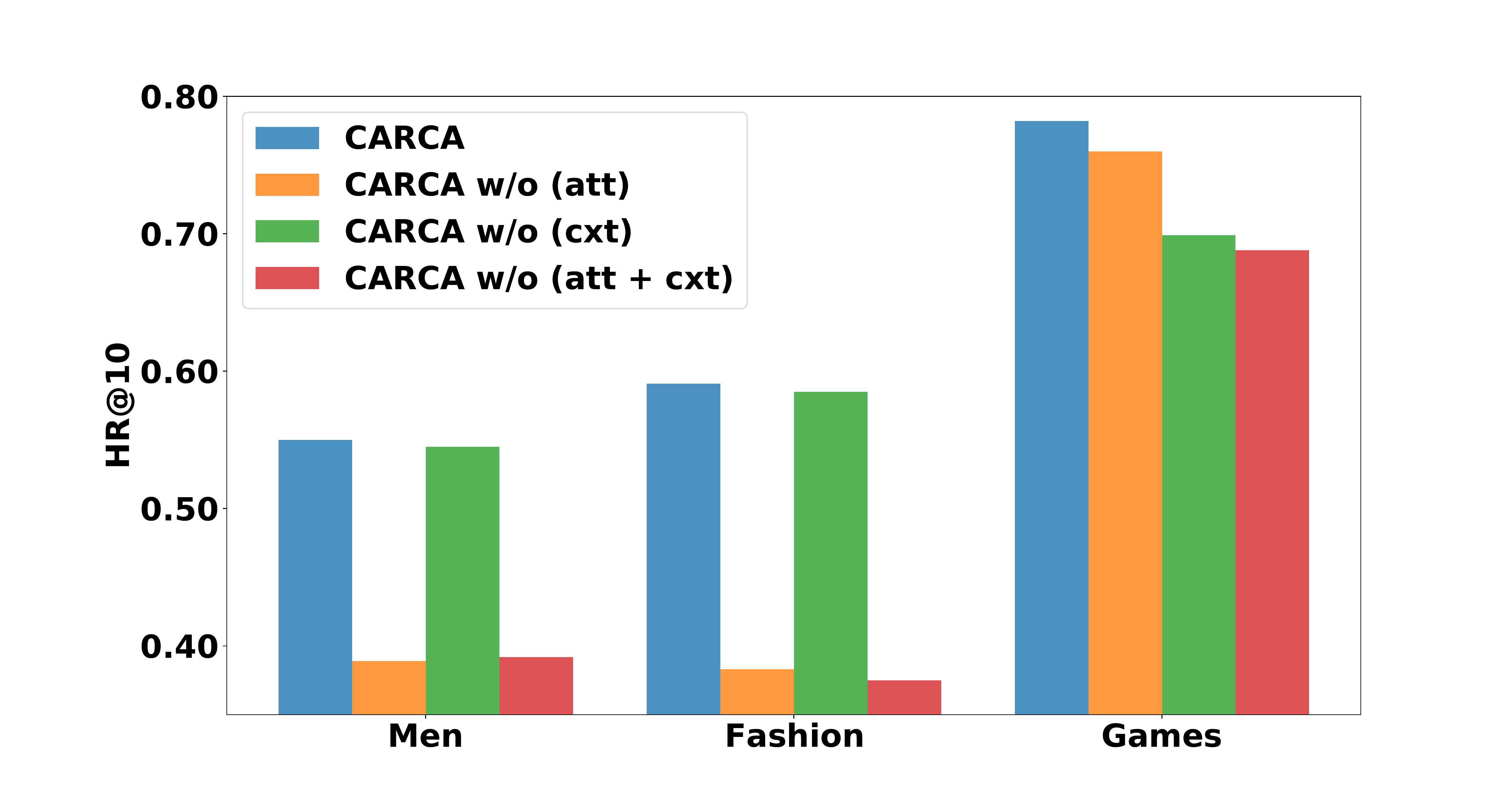}
  \caption{HitRatio}
  \label{sub2}
\end{subfigure}
\caption{Impact of the item's attributes and the contextual features on CARCA's performance.}
\label{feat} 
\end{figure} 
%%%%%%%%%%%%%%%%%%%%%%%%%
\subsection{Impact of item attributes and contextual features (RQ3)}
To measure the impact of the item attributes and the contextual features on CARCA's performance, we conducted a comparative study between different CARCA versions that utilize different combinations of attributes and features. Figure \ref{feat} shows that the impact of item attributes and contextual features are different across the datasets. On the Men and Fashion datasets, item attributes such as their image features have a significant impact on the performance compared to the interactions contexts that had a lower impact. On the other hand, contextual features have a higher impact on CARCA's performance on the Games dataset than item attributes because video games are much more volatile than clothes and fashion-based products as they are susceptible to critics and the satisfaction of their player-bases.

\subsection{Ablation Study (RQ4)}
To measure the impact of each of the model components, multiple experiments on the Men dataset were conducted to compare the different possible configurations of the CARCA architecture. The hyper-parameters of each configuration were optimized separately on the validation set using grid search.
\subsubsection{Configurations}
\begin{enumerate}

 \item \textbf{Default}: This the default configuration of CARCA that was described in Section 4.
  \item \textbf{Additive residual connections }: CARCA with additive residual connections in the multi-head attention blocks instead of multiplicative ones.
  \item \textbf{Concat. all features }: CARCA with all input vectors concatenated in one long feature vector that goes through one embedding layer.
   \item \textbf{Concat. item features }: CARCA with flipped feature extraction pipeline where item attributes are concatenated with the item's one-hot vectors instead of the contextual features.
\item \textbf{Positional Encoding }: CARCA with positional encoding instead of contextual features.
\item \textbf{Additional self-attention blocks on output }: CARCA with an additional multi-head attention block on top of the latent output vectors.
\item \textbf{CARCA with single target split } ; CARCA model trained with only one target item in the target items lists $O^{u(+)} =\{i^{P}_{|P^{u}_{t}|}\}$ and $O^{u(-)} = \{i \notin P^{u}\}$.
\item \textbf{CARCA with transformer architecture }: CARCA with a similar architecture to the original transformers, which is a combination of configurations (2), (3), (5), and (7).

\end{enumerate}
\subsubsection{Results}
Table \ref{Abl} shows that replacing the context features with traditional positional encodings has a slight negative effect on the performance, indicating that the positional encoding can be used as a substitute for contextual features of the interaction if timestamps are missing. Results also show that the multiplicative residual connection in the multi-head attention blocks provides better prediction performance as they can recover the element-wise multiplicative interactions between the latent features similar to the matrix factorization techniques. Additionally, results show that using a single target for training the CARCA model is significantly inferior to using list-wise target items because the model failed to capture the sequential correlation between the inputs and target lists during training as this suppresses its bidirectional auto-regressive capability. However, this can be solved by splitting the whole user's history into sequences with different lengths in a rolling window fashion. Moreover, in Table \ref{Abl}, we can see that the combination strategy of the item attributes and contextual features has different impacts on the overall performance, with the default configuration being the best.
Finally, results also showed that using the transformer architecture or using additional attention blocks on top of the cross-attention block has a significant negative effect on the overall performance.

%%%%%%%%%%%%%%%%%%%%%%%%%%%%%%%%%%%%%%%%%%%%%%%%%%%
\begin{table}[!ht]
\caption{Ablation analysis between different CARCA configurations on the Men dataset.}

\label{Abl}
\small
\begin{center}
\begin{tabular}{lcc}
\toprule

                       Configuration              & HR@10 & NDCG@10   \\
 \midrule                      
Default (1)                                            &\textbf{ 0.550 }& \textbf{0.349} \\
Additive residual connections (2)                  & 0.513 & 0.325 \\
Concat. all features (3)                            & 0.543 & 0.340 \\
Concat. item features (4)     & 0.540 & 0.339 \\
Positional encoding (5)                            & \underline{0.544} & \underline{0.345} \\
Additional self-attention blocks on output (6) &0.427 & 0.231	 \\
CARCA with single target split (7) & 0.394 & 0.233	 \\
CARCA with transformer architecture (8) & 0.459  &  0.276 	 \\
\bottomrule
\end{tabular}
  \end{center}
\end{table}
%%%%%%%%%%%%%%%%%%%%%%%%%%%%%%%%%%%%%%%%%%%%%%%%%%%%
\section{Conclusion and Future Work}

In this paper, we propose CARCA, a context and attribute-aware model that captures user profiles' dynamic nature and contextual changes seamlessly alongside leveraging any available item attributes. CARCA also uses a cross-attention component for scoring items by capturing the correlation between old and recent items in the user profile and their influence on deciding which item to recommend next. Experimental results on four diverse datasets show that CARCA significantly outperforms multiple state-of-the-art models on the task of item recommendation.
%%%%

In future works, we plan to extend CARCA for next-basket recommendations scenarios. We also plan to extend the model capacity by exploring different regularization techniques such as stochastic shared embeddings.


%%%%%%%%%%%%%%%%%%%%%%%%
\begin{table*}[!ht]

\caption{Hyper-parameters configurations and search space of the proposed models.}
\label{hypProp}
\footnotesize
\begin{tabular}{lllllll}
\toprule
\multirow{2}{*}{Model}          & \multirow{2}{*}{param}                                                                      & \multicolumn{4}{c}{Dataset Best Parameters} & \multirow{2}{*}{Search Space}                            \\
                                &                                                                                             & Men        & Fashion   & Games   & Beauty   &                                                          \\
\midrule
\multirow{8}{*}{SASRec++}       & Learning Rate                                                                               & 0.00001    & 0.00001   & 0.0001  & 0.00002  & \{0.001, 0.0001, 0.00002, 0.00001\}                      \\
                                & Max Seq. Length                                                                             & 50         & 35        & 50      & 50       & {[}10, 200{]}, Step size = 5                             \\
                                & Number of attention blocks                                                                  & 3          & 3         & 1       & 3        & \{1, 2, 3, 4, 5\}                                        \\
                                & Number of attention heads                                                                   & 3          & 1         & 1       & 1        & \{1, 2, 3, 4, 5\}                                        \\
                                & Dropout Rate                                                                                & 0.8        & 0.5       & 0.5     & 0.5      & {[}0.0, 1.0{]}, Step size = 0.1                          \\
                                & L2 reg. weight                                                                              & 0          & 0         & 0.001   & 0.001    & {[}0, 0.001,0.0001,0.00001, 0.000001{]}                     \\
                                & Embeding Dimension d                                                                        & 60         & 30        & 50      & 90       & {[}10, 500{]}, Step size = 10                            \\
                                & Embeding Dimension g                                                                        & 300        & 150       & 250     & 450      & {[}50, 500{]}, Step size = 50                            \\
\midrule
\multirow{9}{*}{CARCA}          & Learning Rate                                                                               & 0.000006   & 0.00001   & 0.0001  & 0.0001   & \{0.001, 0.0001, 0.00001, 0.000006, 0.000002, 0.000001\} \\
                                & Max Seq. Length                                                                             & 35         & 35        & 50      & 75       & {[}10, 200{]}, Step size = 5                             \\
                                & Number of attention blocks                                                                  & 3          & 3         & 3       & 3        & \{1, 2, 3, 4, 5\}                                        \\
                                & Number of attention heads                                                                   & 3          & 3         & 3       & 1        & \{1, 2, 3, 4, 5\}                                        \\
                                & Dropout Rate                                                                                & 0.3        & 0.3       & 0.5     & 0.5      & {[}0.0, 1.0{]}, Step size = 0.1                          \\
                                & L2 reg. weight                                                                              & 0.0001     & 0.0001    & 0       & 0.0001   & {[}0, 0.001,0.0001,0.00001, 0.000001{]}                 \\
                                & Embeding Dimension d                                                                        & 390        & 390       & 90      & 90       & {[}10, 500{]}, Step size = 10                            \\
                                & Embeding Dimension g                                                                        & 1950       & 1950      & 450     & 450      & {[}50, 2000{]}, Step size = 50                           \\
                                & \begin{tabular}[c]{@{}l@{}}Residual connection \\ in the cross-attention block\end{tabular} & No         & No        & Yes     & Yes      & \{No, Yes\}      \\
\bottomrule
\end{tabular}
\end{table*}
%%%%%%%%%%%%%%%%%%%%%%%%%%
%%%%%%%%%%%%%%%%%%%%%%%

\begin{table}[!ht]
\caption{Runtime Comparison on Games Dataset}
\label{Runt}
\footnotesize
\begin{tabular}{lc}
\toprule
Model          & Average batch runtime in seconds \\
\midrule
S\textsuperscript{3}Rec      & 0.580                             \\
SSE-PT         & 0.008                            \\
SASRec        & 0.013                            \\
TiSASRec        & 0.075                          \\
SSE-SASRec     & 0.015                            \\
OAR            & 0.018                            \\
\midrule
SASRec++       & 0.028                            \\
\midrule
CARCA (w/o CA) & 0.015                            \\
CARCA          & 0.026                           \\
\bottomrule
\end{tabular}
\end{table}
%%%%%%%%%%%%%%%%%%%%%%%%%

\appendix
\section{Runtime Comparison}

We compared the average run-time of a batch of size 128 between the proposed CARCA models and the most recent state-of-the-art sequential baseline models on the Games dataset using 1080Ti GPU, and E5-1660v4 CPU with 64GB of RAM. Results in table \ref{Runt} show that the average batch run-time of CARCA is very close to other sequential models with very little overhead due to adding the cross-attention component and the additional features. Results also show that CARCA is about \textbf{22 times} faster than S\textsuperscript{3}Rec, which is the closest competitor because of S\textsuperscript{3}Rec's computationally expensive mutual information maximization procedure.

%%%%%%%%%%%%%%%%%%%%%%%%%%%

%%%%%%%%%%%%%%%%%%%%

\section{Reproducibility of the Experiments}
The source code of CARCA\footnote{\url{https://github.com/ahmedrashed-ml/CARCA}} is available at github. Regarding the hyper-parameters, all models were tuned using a grid search except the EASE model as it was optimized using a Bayesian optimization procedure existed in its source code\footnote{https://github.com/MaurizioFD/RecSys2019\_DeepLearning\_Evaluation} by Dacrema et al. \cite{dacrema2019we}. The best found hyper-parmeters for our CARCA models is shown in Table \ref{hypProp}

The source codes of all baseline models are available at their authors' GitHub repositories  \footnote{https://github.com/ahmedrashed-ml/GraphRec}\footnote{https://github.com/ChenglongChen/tensorflow-DeepFM}\footnote{https://github.com/kang205/SASRec}\footnote{https://github.com/wangjlgz/Occasion-Aware-Recommenation}\footnote{https://github.com/wuliwei9278/SSE-PT/tree/master/SSE-SASrec}\footnote{https://github.com/wuliwei9278/SSE-PT}\footnote{https://github.com/RUCAIBox/CIKM2020-S3Rec}\footnote{https://github.com/FeiSun/BERT4Rec}\footnote{https://github.com/JiachengLi1995/TiSASRec}.

\bibliographystyle{ACM-Reference-Format}
\bibliography{sample-base}


\begin{thebibliography}{36}

%%% the Web address.

\ifx \showCODEN    \undefined \def \showCODEN     #1{\unskip}     \fi
\ifx \showDOI      \undefined \def \showDOI       #1{#1}\fi
\ifx \showISBNx    \undefined \def \showISBNx     #1{\unskip}     \fi
\ifx \showISBNxiii \undefined \def \showISBNxiii  #1{\unskip}     \fi
\ifx \showISSN     \undefined \def \showISSN      #1{\unskip}     \fi
\ifx \showLCCN     \undefined \def \showLCCN      #1{\unskip}     \fi
\ifx \shownote     \undefined \def \shownote      #1{#1}          \fi
\ifx \showarticletitle \undefined \def \showarticletitle #1{#1}   \fi
\ifx \showURL      \undefined \def \showURL       {\relax}        \fi
% invisible to TeX
\providecommand\bibfield[2]{#2}
\providecommand\bibinfo[2]{#2}
\providecommand\natexlab[1]{#1}
\providecommand\showeprint[2][]{arXiv:#2}

\bibitem[\protect\citeauthoryear{Ba, Kiros, and Hinton}{Ba
  et~al\mbox{.}}{2016}]%
        {ba2016layer}
\bibfield{author}{\bibinfo{person}{Jimmy~Lei Ba}, \bibinfo{person}{Jamie~Ryan
  Kiros}, {and} \bibinfo{person}{Geoffrey~E Hinton}.}
  \bibinfo{year}{2016}\natexlab{}.
\newblock \showarticletitle{Layer Normalization}.
\newblock \bibinfo{journal}{\emph{stat}}  \bibinfo{volume}{1050}
  (\bibinfo{year}{2016}), \bibinfo{pages}{21}.
\newblock


\bibitem[\protect\citeauthoryear{Dacrema, Cremonesi, and Jannach}{Dacrema
  et~al\mbox{.}}{2019}]%
        {dacrema2019we}
\bibfield{author}{\bibinfo{person}{Maurizio~Ferrari Dacrema},
  \bibinfo{person}{Paolo Cremonesi}, {and} \bibinfo{person}{Dietmar Jannach}.}
  \bibinfo{year}{2019}\natexlab{}.
\newblock \showarticletitle{Are we really making much progress? A worrying
  analysis of recent neural recommendation approaches}. In
  \bibinfo{booktitle}{\emph{Proceedings of the 13th ACM Conference on
  Recommender Systems}}. \bibinfo{pages}{101--109}.
\newblock


\bibitem[\protect\citeauthoryear{Deng, Dong, Socher, Li, Li, and Fei-Fei}{Deng
  et~al\mbox{.}}{2009}]%
        {imagenet_cvpr09}
\bibfield{author}{\bibinfo{person}{J. Deng}, \bibinfo{person}{W. Dong},
  \bibinfo{person}{R. Socher}, \bibinfo{person}{L.-J. Li}, \bibinfo{person}{K.
  Li}, {and} \bibinfo{person}{L. Fei-Fei}.} \bibinfo{year}{2009}\natexlab{}.
\newblock \showarticletitle{{ImageNet: A Large-Scale Hierarchical Image
  Database}}. In \bibinfo{booktitle}{\emph{CVPR09}}.
\newblock


\bibitem[\protect\citeauthoryear{Devlin, Chang, Lee, and Toutanova}{Devlin
  et~al\mbox{.}}{2019}]%
        {devlin2019bert}
\bibfield{author}{\bibinfo{person}{Jacob Devlin}, \bibinfo{person}{Ming-Wei
  Chang}, \bibinfo{person}{Kenton Lee}, {and} \bibinfo{person}{Kristina
  Toutanova}.} \bibinfo{year}{2019}\natexlab{}.
\newblock \showarticletitle{BERT: Pre-training of Deep Bidirectional
  Transformers for Language Understanding}. In
  \bibinfo{booktitle}{\emph{Proceedings of the 2019 Conference of the North
  American Chapter of the Association for Computational Linguistics: Human
  Language Technologies, Volume 1 (Long and Short Papers)}}.
  \bibinfo{pages}{4171--4186}.
\newblock


\bibitem[\protect\citeauthoryear{Ferrari~Dacrema, Parroni, Cremonesi, and
  Jannach}{Ferrari~Dacrema et~al\mbox{.}}{2020}]%
        {ferrari2020critically}
\bibfield{author}{\bibinfo{person}{Maurizio Ferrari~Dacrema},
  \bibinfo{person}{Federico Parroni}, \bibinfo{person}{Paolo Cremonesi}, {and}
  \bibinfo{person}{Dietmar Jannach}.} \bibinfo{year}{2020}\natexlab{}.
\newblock \showarticletitle{Critically Examining the Claimed Value of
  Convolutions over User-Item Embedding Maps for Recommender Systems}. In
  \bibinfo{booktitle}{\emph{Proceedings of the 29th ACM International
  Conference on Information \& Knowledge Management}}.
  \bibinfo{pages}{355--363}.
\newblock


\bibitem[\protect\citeauthoryear{Guo, Tang, Ye, Li, and He}{Guo
  et~al\mbox{.}}{2017}]%
        {guo2017deepfm}
\bibfield{author}{\bibinfo{person}{Huifeng Guo}, \bibinfo{person}{Ruiming
  Tang}, \bibinfo{person}{Yunming Ye}, \bibinfo{person}{Zhenguo Li}, {and}
  \bibinfo{person}{Xiuqiang He}.} \bibinfo{year}{2017}\natexlab{}.
\newblock \showarticletitle{DeepFM: A Factorization-Machine based Neural
  Network for CTR Prediction}. In \bibinfo{booktitle}{\emph{IJCAI}}.
\newblock


\bibitem[\protect\citeauthoryear{He, Zhang, Ren, and Sun}{He
  et~al\mbox{.}}{2016}]%
        {he2016deep}
\bibfield{author}{\bibinfo{person}{Kaiming He}, \bibinfo{person}{Xiangyu
  Zhang}, \bibinfo{person}{Shaoqing Ren}, {and} \bibinfo{person}{Jian Sun}.}
  \bibinfo{year}{2016}\natexlab{}.
\newblock \showarticletitle{Deep residual learning for image recognition}. In
  \bibinfo{booktitle}{\emph{Proceedings of the IEEE conference on computer
  vision and pattern recognition}}. \bibinfo{pages}{770--778}.
\newblock


\bibitem[\protect\citeauthoryear{He and McAuley}{He and McAuley}{2016}]%
        {he2016vbpr}
\bibfield{author}{\bibinfo{person}{Ruining He} {and} \bibinfo{person}{Julian
  McAuley}.} \bibinfo{year}{2016}\natexlab{}.
\newblock \showarticletitle{VBPR: visual bayesian personalized ranking from
  implicit feedback}. In \bibinfo{booktitle}{\emph{Proceedings of the AAAI
  Conference on Artificial Intelligence}}, Vol.~\bibinfo{volume}{30}.
\newblock


\bibitem[\protect\citeauthoryear{He and Chua}{He and Chua}{2017}]%
        {he2017neural}
\bibfield{author}{\bibinfo{person}{Xiangnan He} {and} \bibinfo{person}{Tat-Seng
  Chua}.} \bibinfo{year}{2017}\natexlab{}.
\newblock \showarticletitle{Neural factorization machines for sparse predictive
  analytics}. In \bibinfo{booktitle}{\emph{Proceedings of the 40th
  International ACM SIGIR conference on Research and Development in Information
  Retrieval}}. \bibinfo{pages}{355--364}.
\newblock


\bibitem[\protect\citeauthoryear{Hidasi and Karatzoglou}{Hidasi and
  Karatzoglou}{2018}]%
        {hidasi2018recurrent}
\bibfield{author}{\bibinfo{person}{Bal{\'a}zs Hidasi} {and}
  \bibinfo{person}{Alexandros Karatzoglou}.} \bibinfo{year}{2018}\natexlab{}.
\newblock \showarticletitle{Recurrent neural networks with top-k gains for
  session-based recommendations}. In \bibinfo{booktitle}{\emph{Proceedings of
  the 27th ACM international conference on information and knowledge
  management}}. \bibinfo{pages}{843--852}.
\newblock


\bibitem[\protect\citeauthoryear{Hou, Wu, Chen, Li, Zheng, and Liu}{Hou
  et~al\mbox{.}}{2019}]%
        {ijcai2019650}
\bibfield{author}{\bibinfo{person}{Min Hou}, \bibinfo{person}{Le Wu},
  \bibinfo{person}{Enhong Chen}, \bibinfo{person}{Zhi Li},
  \bibinfo{person}{Vincent~W. Zheng}, {and} \bibinfo{person}{Qi Liu}.}
  \bibinfo{year}{2019}\natexlab{}.
\newblock \showarticletitle{Explainable Fashion Recommendation: A Semantic
  Attribute Region Guided Approach}. In \bibinfo{booktitle}{\emph{Proceedings
  of the Twenty-Eighth International Joint Conference on Artificial
  Intelligence, {IJCAI-19}}}. \bibinfo{publisher}{International Joint
  Conferences on Artificial Intelligence Organization},
  \bibinfo{pages}{4681--4688}.
\newblock
\urldef\tempurl%
\url{https://doi.org/10.24963/ijcai.2019/650}
\showDOI{\tempurl}


\bibitem[\protect\citeauthoryear{Kang and McAuley}{Kang and McAuley}{2018}]%
        {kang2018self}
\bibfield{author}{\bibinfo{person}{Wang-Cheng Kang} {and}
  \bibinfo{person}{Julian McAuley}.} \bibinfo{year}{2018}\natexlab{}.
\newblock \showarticletitle{Self-attentive sequential recommendation}. In
  \bibinfo{booktitle}{\emph{2018 IEEE International Conference on Data Mining
  (ICDM)}}. IEEE, \bibinfo{pages}{197--206}.
\newblock


\bibitem[\protect\citeauthoryear{Li, Wang, and McAuley}{Li
  et~al\mbox{.}}{2020}]%
        {li2020time}
\bibfield{author}{\bibinfo{person}{Jiacheng Li}, \bibinfo{person}{Yujie Wang},
  {and} \bibinfo{person}{Julian McAuley}.} \bibinfo{year}{2020}\natexlab{}.
\newblock \showarticletitle{Time interval aware self-attention for sequential
  recommendation}. In \bibinfo{booktitle}{\emph{Proceedings of the 13th
  international conference on web search and data mining}}.
  \bibinfo{pages}{322--330}.
\newblock


\bibitem[\protect\citeauthoryear{Li, Kawale, and Fu}{Li et~al\mbox{.}}{2015}]%
        {li2015deep}
\bibfield{author}{\bibinfo{person}{Sheng Li}, \bibinfo{person}{Jaya Kawale},
  {and} \bibinfo{person}{Yun Fu}.} \bibinfo{year}{2015}\natexlab{}.
\newblock \showarticletitle{Deep collaborative filtering via marginalized
  denoising auto-encoder}. In \bibinfo{booktitle}{\emph{Proceedings of the 24th
  ACM international on conference on information and knowledge management}}.
  \bibinfo{pages}{811--820}.
\newblock


\bibitem[\protect\citeauthoryear{Ma, Kang, and Liu}{Ma et~al\mbox{.}}{2019}]%
        {ma2019hierarchical}
\bibfield{author}{\bibinfo{person}{Chen Ma}, \bibinfo{person}{Peng Kang}, {and}
  \bibinfo{person}{Xue Liu}.} \bibinfo{year}{2019}\natexlab{}.
\newblock \showarticletitle{Hierarchical gating networks for sequential
  recommendation}. In \bibinfo{booktitle}{\emph{Proceedings of the 25th ACM
  SIGKDD International Conference on Knowledge Discovery \& Data Mining}}.
  \bibinfo{pages}{825--833}.
\newblock


\bibitem[\protect\citeauthoryear{Rashed, Grabocka, and Schmidt-Thieme}{Rashed
  et~al\mbox{.}}{2019}]%
        {rashed2019attribute}
\bibfield{author}{\bibinfo{person}{Ahmed Rashed}, \bibinfo{person}{Josif
  Grabocka}, {and} \bibinfo{person}{Lars Schmidt-Thieme}.}
  \bibinfo{year}{2019}\natexlab{}.
\newblock \showarticletitle{Attribute-aware non-linear co-embeddings of graph
  features}. In \bibinfo{booktitle}{\emph{Proceedings of the 13th ACM
  Conference on Recommender Systems}}. \bibinfo{pages}{314--321}.
\newblock


\bibitem[\protect\citeauthoryear{Rashed, Jawed, Schmidt-Thieme, and
  Hintsches}{Rashed et~al\mbox{.}}{2020}]%
        {rashed2020multirec}
\bibfield{author}{\bibinfo{person}{Ahmed Rashed}, \bibinfo{person}{Shayan
  Jawed}, \bibinfo{person}{Lars Schmidt-Thieme}, {and} \bibinfo{person}{Andre
  Hintsches}.} \bibinfo{year}{2020}\natexlab{}.
\newblock \showarticletitle{MultiRec: A Multi-Relational Approach for Unique
  Item Recommendation in Auction Systems}. In
  \bibinfo{booktitle}{\emph{Fourteenth ACM Conference on Recommender Systems}}.
  \bibinfo{pages}{230--239}.
\newblock


\bibitem[\protect\citeauthoryear{Rendle}{Rendle}{2010}]%
        {rendle2010factorization}
\bibfield{author}{\bibinfo{person}{Steffen Rendle}.}
  \bibinfo{year}{2010}\natexlab{}.
\newblock \showarticletitle{Factorization machines}. In
  \bibinfo{booktitle}{\emph{2010 IEEE International Conference on Data
  Mining}}. IEEE, \bibinfo{pages}{995--1000}.
\newblock


\bibitem[\protect\citeauthoryear{Rendle, Freudenthaler, Gantner, and
  Schmidt-Thieme}{Rendle et~al\mbox{.}}{2012}]%
        {rendle2012bpr}
\bibfield{author}{\bibinfo{person}{Steffen Rendle}, \bibinfo{person}{Christoph
  Freudenthaler}, \bibinfo{person}{Zeno Gantner}, {and} \bibinfo{person}{Lars
  Schmidt-Thieme}.} \bibinfo{year}{2012}\natexlab{}.
\newblock \showarticletitle{BPR: Bayesian personalized ranking from implicit
  feedback}.
\newblock \bibinfo{journal}{\emph{arXiv preprint arXiv:1205.2618}}
  (\bibinfo{year}{2012}).
\newblock


\bibitem[\protect\citeauthoryear{Song, Shi, Xiao, Duan, Xu, Zhang, and
  Tang}{Song et~al\mbox{.}}{2019}]%
        {song2019autoint}
\bibfield{author}{\bibinfo{person}{Weiping Song}, \bibinfo{person}{Chence Shi},
  \bibinfo{person}{Zhiping Xiao}, \bibinfo{person}{Zhijian Duan},
  \bibinfo{person}{Yewen Xu}, \bibinfo{person}{Ming Zhang}, {and}
  \bibinfo{person}{Jian Tang}.} \bibinfo{year}{2019}\natexlab{}.
\newblock \showarticletitle{Autoint: Automatic feature interaction learning via
  self-attentive neural networks}. In \bibinfo{booktitle}{\emph{Proceedings of
  the 28th ACM International Conference on Information and Knowledge
  Management}}. \bibinfo{pages}{1161--1170}.
\newblock


\bibitem[\protect\citeauthoryear{Srivastava, Hinton, Krizhevsky, Sutskever, and
  Salakhutdinov}{Srivastava et~al\mbox{.}}{2014}]%
        {srivastava2014dropout}
\bibfield{author}{\bibinfo{person}{Nitish Srivastava},
  \bibinfo{person}{Geoffrey Hinton}, \bibinfo{person}{Alex Krizhevsky},
  \bibinfo{person}{Ilya Sutskever}, {and} \bibinfo{person}{Ruslan
  Salakhutdinov}.} \bibinfo{year}{2014}\natexlab{}.
\newblock \showarticletitle{Dropout: a simple way to prevent neural networks
  from overfitting}.
\newblock \bibinfo{journal}{\emph{The journal of machine learning research}}
  \bibinfo{volume}{15}, \bibinfo{number}{1} (\bibinfo{year}{2014}),
  \bibinfo{pages}{1929--1958}.
\newblock


\bibitem[\protect\citeauthoryear{Steck}{Steck}{2019}]%
        {steck2019embarrassingly}
\bibfield{author}{\bibinfo{person}{Harald Steck}.}
  \bibinfo{year}{2019}\natexlab{}.
\newblock \showarticletitle{Embarrassingly shallow autoencoders for sparse
  data}. In \bibinfo{booktitle}{\emph{The World Wide Web Conference}}.
  \bibinfo{pages}{3251--3257}.
\newblock


\bibitem[\protect\citeauthoryear{Sun, Liu, Wu, Pei, Lin, Ou, and Jiang}{Sun
  et~al\mbox{.}}{2019}]%
        {sun2019bert4rec}
\bibfield{author}{\bibinfo{person}{Fei Sun}, \bibinfo{person}{Jun Liu},
  \bibinfo{person}{Jian Wu}, \bibinfo{person}{Changhua Pei},
  \bibinfo{person}{Xiao Lin}, \bibinfo{person}{Wenwu Ou}, {and}
  \bibinfo{person}{Peng Jiang}.} \bibinfo{year}{2019}\natexlab{}.
\newblock \showarticletitle{BERT4Rec: Sequential recommendation with
  bidirectional encoder representations from transformer}. In
  \bibinfo{booktitle}{\emph{Proceedings of the 28th ACM international
  conference on information and knowledge management}}.
  \bibinfo{pages}{1441--1450}.
\newblock


\bibitem[\protect\citeauthoryear{Vaswani, Shazeer, Parmar, Uszkoreit, Jones,
  Gomez, Kaiser, and Polosukhin}{Vaswani et~al\mbox{.}}{2017}]%
        {vaswani2017attention}
\bibfield{author}{\bibinfo{person}{Ashish Vaswani}, \bibinfo{person}{Noam
  Shazeer}, \bibinfo{person}{Niki Parmar}, \bibinfo{person}{Jakob Uszkoreit},
  \bibinfo{person}{Llion Jones}, \bibinfo{person}{Aidan~N Gomez},
  \bibinfo{person}{Lukasz Kaiser}, {and} \bibinfo{person}{Illia Polosukhin}.}
  \bibinfo{year}{2017}\natexlab{}.
\newblock \showarticletitle{Attention is all you need}.
\newblock \bibinfo{journal}{\emph{arXiv preprint arXiv:1706.03762}}
  (\bibinfo{year}{2017}).
\newblock


\bibitem[\protect\citeauthoryear{Wang, Ding, Hong, Liu, and Caverlee}{Wang
  et~al\mbox{.}}{2020a}]%
        {wang2020next}
\bibfield{author}{\bibinfo{person}{Jianling Wang}, \bibinfo{person}{Kaize
  Ding}, \bibinfo{person}{Liangjie Hong}, \bibinfo{person}{Huan Liu}, {and}
  \bibinfo{person}{James Caverlee}.} \bibinfo{year}{2020}\natexlab{a}.
\newblock \showarticletitle{Next-item recommendation with sequential
  hypergraphs}. In \bibinfo{booktitle}{\emph{Proceedings of the 43rd
  International ACM SIGIR Conference on Research and Development in Information
  Retrieval}}. \bibinfo{pages}{1101--1110}.
\newblock


\bibitem[\protect\citeauthoryear{Wang, Louca, Hu, Cellier, Caverlee, and
  Hong}{Wang et~al\mbox{.}}{2020b}]%
        {wang2020time}
\bibfield{author}{\bibinfo{person}{Jianling Wang}, \bibinfo{person}{Raphael
  Louca}, \bibinfo{person}{Diane Hu}, \bibinfo{person}{Caitlin Cellier},
  \bibinfo{person}{James Caverlee}, {and} \bibinfo{person}{Liangjie Hong}.}
  \bibinfo{year}{2020}\natexlab{b}.
\newblock \showarticletitle{Time to Shop for Valentine's Day: Shopping
  Occasions and Sequential Recommendation in E-commerce}. In
  \bibinfo{booktitle}{\emph{Proceedings of the 13th International Conference on
  Web Search and Data Mining}}. \bibinfo{pages}{645--653}.
\newblock


\bibitem[\protect\citeauthoryear{Wu, Li, Hsieh, and Sharpnack}{Wu
  et~al\mbox{.}}{2019}]%
        {wu2019stochastic}
\bibfield{author}{\bibinfo{person}{Liwei Wu}, \bibinfo{person}{Shuqing Li},
  \bibinfo{person}{Cho-Jui Hsieh}, {and} \bibinfo{person}{James Sharpnack}.}
  \bibinfo{year}{2019}\natexlab{}.
\newblock \showarticletitle{Stochastic Shared Embeddings: Data-driven
  Regularization of Embedding Layers}.
\newblock \bibinfo{journal}{\emph{NeurIPS}} (\bibinfo{year}{2019}).
\newblock


\bibitem[\protect\citeauthoryear{Wu, Li, Hsieh, and Sharpnack}{Wu
  et~al\mbox{.}}{2020}]%
        {wu2020sse}
\bibfield{author}{\bibinfo{person}{Liwei Wu}, \bibinfo{person}{Shuqing Li},
  \bibinfo{person}{Cho-Jui Hsieh}, {and} \bibinfo{person}{James Sharpnack}.}
  \bibinfo{year}{2020}\natexlab{}.
\newblock \showarticletitle{SSE-PT: Sequential recommendation via personalized
  transformer}. In \bibinfo{booktitle}{\emph{Fourteenth ACM Conference on
  Recommender Systems}}. \bibinfo{pages}{328--337}.
\newblock


\bibitem[\protect\citeauthoryear{Xiao, Ye, He, Zhang, Wu, and Chua}{Xiao
  et~al\mbox{.}}{2017}]%
        {xiao2017attentional}
\bibfield{author}{\bibinfo{person}{Jun Xiao}, \bibinfo{person}{Hao Ye},
  \bibinfo{person}{Xiangnan He}, \bibinfo{person}{Hanwang Zhang},
  \bibinfo{person}{Fei Wu}, {and} \bibinfo{person}{Tat-Seng Chua}.}
  \bibinfo{year}{2017}\natexlab{}.
\newblock \showarticletitle{Attentional factorization machines: learning the
  weight of feature interactions via attention networks}. In
  \bibinfo{booktitle}{\emph{Proceedings of the 26th International Joint
  Conference on Artificial Intelligence}}. \bibinfo{pages}{3119--3125}.
\newblock


\bibitem[\protect\citeauthoryear{Xin, Chen, He, Wang, Ding, and Jose}{Xin
  et~al\mbox{.}}{2019}]%
        {xin2019cfm}
\bibfield{author}{\bibinfo{person}{Xin Xin}, \bibinfo{person}{Bo Chen},
  \bibinfo{person}{Xiangnan He}, \bibinfo{person}{Dong Wang},
  \bibinfo{person}{Yue Ding}, {and} \bibinfo{person}{Joemon Jose}.}
  \bibinfo{year}{2019}\natexlab{}.
\newblock \showarticletitle{CFM: Convolutional Factorization Machines for
  Context-Aware Recommendation.}. In \bibinfo{booktitle}{\emph{IJCAI}},
  Vol.~\bibinfo{volume}{19}. \bibinfo{pages}{3926--3932}.
\newblock


\bibitem[\protect\citeauthoryear{Zhang, Yao, and Xu}{Zhang
  et~al\mbox{.}}{2017b}]%
        {zhang2017autosvd++}
\bibfield{author}{\bibinfo{person}{Shuai Zhang}, \bibinfo{person}{Lina Yao},
  {and} \bibinfo{person}{Xiwei Xu}.} \bibinfo{year}{2017}\natexlab{b}.
\newblock \showarticletitle{AutoSVD++ An Efficient Hybrid Collaborative
  Filtering Model via Contractive Auto-encoders}. In
  \bibinfo{booktitle}{\emph{Proceedings of the 40th International ACM SIGIR
  conference on Research and Development in Information Retrieval}}.
  \bibinfo{pages}{957--960}.
\newblock


\bibitem[\protect\citeauthoryear{Zhang, Zhao, Liu, Sheng, Xu, Wang, Liu, and
  Zhou}{Zhang et~al\mbox{.}}{2019}]%
        {zhang2019feature}
\bibfield{author}{\bibinfo{person}{Tingting Zhang}, \bibinfo{person}{Pengpeng
  Zhao}, \bibinfo{person}{Yanchi Liu}, \bibinfo{person}{Victor~S Sheng},
  \bibinfo{person}{Jiajie Xu}, \bibinfo{person}{Deqing Wang},
  \bibinfo{person}{Guanfeng Liu}, {and} \bibinfo{person}{Xiaofang Zhou}.}
  \bibinfo{year}{2019}\natexlab{}.
\newblock \showarticletitle{Feature-level Deeper Self-Attention Network for
  Sequential Recommendation.}. In \bibinfo{booktitle}{\emph{IJCAI}}.
  \bibinfo{pages}{4320--4326}.
\newblock


\bibitem[\protect\citeauthoryear{Zhang, Ai, Chen, and Croft}{Zhang
  et~al\mbox{.}}{2017a}]%
        {zhang2017joint}
\bibfield{author}{\bibinfo{person}{Yongfeng Zhang}, \bibinfo{person}{Qingyao
  Ai}, \bibinfo{person}{Xu Chen}, {and} \bibinfo{person}{W~Bruce Croft}.}
  \bibinfo{year}{2017}\natexlab{a}.
\newblock \showarticletitle{Joint representation learning for top-n
  recommendation with heterogeneous information sources}. In
  \bibinfo{booktitle}{\emph{Proceedings of the 2017 ACM on Conference on
  Information and Knowledge Management}}. \bibinfo{pages}{1449--1458}.
\newblock


\bibitem[\protect\citeauthoryear{Zhou, Mou, Fan, Pi, Bian, Zhou, Zhu, and
  Gai}{Zhou et~al\mbox{.}}{2019}]%
        {zhou2019deep}
\bibfield{author}{\bibinfo{person}{Guorui Zhou}, \bibinfo{person}{Na Mou},
  \bibinfo{person}{Ying Fan}, \bibinfo{person}{Qi Pi}, \bibinfo{person}{Weijie
  Bian}, \bibinfo{person}{Chang Zhou}, \bibinfo{person}{Xiaoqiang Zhu}, {and}
  \bibinfo{person}{Kun Gai}.} \bibinfo{year}{2019}\natexlab{}.
\newblock \showarticletitle{Deep interest evolution network for click-through
  rate prediction}. In \bibinfo{booktitle}{\emph{Proceedings of the AAAI
  conference on artificial intelligence}}, Vol.~\bibinfo{volume}{33}.
  \bibinfo{pages}{5941--5948}.
\newblock


\bibitem[\protect\citeauthoryear{Zhou, Zhu, Song, Fan, Zhu, Ma, Yan, Jin, Li,
  and Gai}{Zhou et~al\mbox{.}}{2018}]%
        {zhou2018deep}
\bibfield{author}{\bibinfo{person}{Guorui Zhou}, \bibinfo{person}{Xiaoqiang
  Zhu}, \bibinfo{person}{Chenru Song}, \bibinfo{person}{Ying Fan},
  \bibinfo{person}{Han Zhu}, \bibinfo{person}{Xiao Ma},
  \bibinfo{person}{Yanghui Yan}, \bibinfo{person}{Junqi Jin},
  \bibinfo{person}{Han Li}, {and} \bibinfo{person}{Kun Gai}.}
  \bibinfo{year}{2018}\natexlab{}.
\newblock \showarticletitle{Deep interest network for click-through rate
  prediction}. In \bibinfo{booktitle}{\emph{Proceedings of the 24th ACM SIGKDD
  International Conference on Knowledge Discovery \& Data Mining}}.
  \bibinfo{pages}{1059--1068}.
\newblock


\bibitem[\protect\citeauthoryear{Zhou, Wang, Zhao, Zhu, Wang, Zhang, Wang, and
  Wen}{Zhou et~al\mbox{.}}{2020}]%
        {ZhouWZZWZWW20}
\bibfield{author}{\bibinfo{person}{Kun Zhou}, \bibinfo{person}{Hui Wang},
  \bibinfo{person}{Wayne~Xin Zhao}, \bibinfo{person}{Yutao Zhu},
  \bibinfo{person}{Sirui Wang}, \bibinfo{person}{Fuzheng Zhang},
  \bibinfo{person}{Zhongyuan Wang}, {and} \bibinfo{person}{Ji{-}Rong Wen}.}
  \bibinfo{year}{2020}\natexlab{}.
\newblock \showarticletitle{S3-Rec: Self-Supervised Learning for Sequential
  Recommendation with Mutual Information Maximization}. In
  \bibinfo{booktitle}{\emph{{CIKM} '20: The 29th {ACM} International Conference
  on Information and Knowledge Management, Virtual Event, Ireland, October
  19-23, 2020}}. \bibinfo{publisher}{{ACM}}, \bibinfo{pages}{1893--1902}.
\newblock


\end{thebibliography}


%%%%%%%%%%%%%%%%%%%%%%%%%%%%%%%%%%%%

%\section{Run-time Comparison}

%%%%%%%%%%%%%%%%%%%%%%%%%%%%%%%%%%%%

\end{document}